\documentclass[12pt,preprint]{aastex}

\slugcomment{RESCEU-6/01 UTAP-389}
\shorttitle{Probing the Core Structure of Dark Halos}
\shortauthors{Oguri, Taruya \& Suto}
\begin{document}
%
\title{Probing the Core Structure of Dark Halos \\
with Tangential and Radial Arc Statistics
}
%
\author{Masamune Oguri, Atsushi Taruya\altaffilmark{1} and Yasushi Suto\altaffilmark{1}}
\affil{Department of Physics, School of Science, University of
    Tokyo, Tokyo 113-0033, Japan.}
\email{oguri@utap.phys.s.u-tokyo.ac.jp,
ataruya@utap.phys.s.u-tokyo.ac.jp, suto@phys.s.u-tokyo.ac.jp}
\altaffiltext{1}{Also at Research Center for the Early Universe (RESCEU), 
School of Science, University of Tokyo, Tokyo 113-0033, Japan.}

%
\received{2001 May 15}
\accepted{2001 ???}
\begin{abstract}
We study the arc statistics of gravitational lensing generated by dark
 matter halos in order to probe their density profile.  We
 characterize the halo profile by two parameters, the inner slope of the
 central cusp $\alpha$, and the median amplitude of the concentration
 parameter, $c_{\rm norm}$, for a halo of mass $10^{14}h^{-1}M_\odot$ at
 $z=0$, and compute the numbers of tangential and radial arcs produced
 by gravitational lensing of galaxy clusters.  We find that the number
 of arcs divided by the number of halos is a good statistic which is
 sensitive to both $c_{\rm norm}$ and $\alpha$ with very weak dependence
 on the cosmological parameters. If the arc samples with well-defined
 selection criteria for the clusters become available, one can strongly
 constrain both $c_{\rm norm}$ and $\alpha$. While our tentative
 comparison with the existing observational data indicates that the
 inner density profile of dark halos is indeed as steep as predicted 
 by recent simulations ($\alpha \sim 1.5$), the
 homogeneous samples of tangential and radial arcs are required for more
 quantitative discussions.
\end{abstract} 
%
%
\section{Introduction}
\label{sec:intro}

Dark matter halos play a central role in the standard picture of the
cosmological structure formation as plausible sites hosting a variety of
astronomical objects such as galaxies and clusters of galaxies.  On the
basis of a series of systematic cosmological N-body simulations,
\citet[hereafter NFW]{navarro96,navarro97} found that the density
profile obeys the ``universal" form $\rho(r)\propto r^{-1}(r+r_{\rm
s})^{-2}$ irrespective of the underlying cosmological parameters, the
shape of the primordial fluctuation spectrum and the formation
histories.  More recent high-resolution simulations indeed confirmed the
existence of the central cusp but suggested the even steeper inner
slope; $\rho(r)\propto r^{-1.5}$ rather than $\propto r^{-1}$
\citep{moore99,fukushige01}. Nevertheless the universality of the
profiles in numerical simulations is fairly established except for
the possible weak dependence on the halo mass \citep{jing00a} and also
for some scatter around the mean \citep{jing00b}.

The above indications from simulations, however, do not seem to be
supported by either simple theoretical considerations or available
observations. Plausible theoretical models rather predict that the inner
slope of the halo profile should depend on the primordial
fluctuation spectrum \citep{hoffman85,syer98} and also on the merging
history (Nusser \& Sheth 1999; see also \L okas \& Hoffman 2000).
Detailed analyses of the X-ray surface
brightness of clusters of galaxies \citep{wu00,wu01} are inconsistent
with the scaling of the halo concentration against the halo mass
predicted by simulations.  Furthermore both rotation curves of the low
surface brightness galaxies \citep{salucci00,deBlok01} and the
inner region of the cluster CL0024-1654 reconstructed from gravitational
lensing images \citep{tyson98} indicate the flat core instead of the
central cusp, although some controversy about each claim still remains
\citep{vandenbosch00,broadhurst00,shapiro00}. This conflict has motivated
wild proposals, including an idea that dark matter is self-interacting
\citep{spergel00}.

Since the current situation concerning the numerical, theoretical and
observational indications for dark matter halo profiles is somewhat
puzzling, it is important to develop another independent methodology to
probe the profiles. For this purpose, we focus on the arc statistics of
the gravitational lensing in the present paper. The major advantages of
this methodology include, (i) the gravitational lensing offers us the
direct route to the mass distribution of the dark halo without
additional assumptions, for instance, on the physical conditions of gas
and stars, (ii) the gravitational arcs are produced mainly due to galaxy
clusters which have an empirically good one-to-one correspondence with
dark halos, in marked contrast with the case of the multiple QSO images
due to galaxies \citep{nakamura97,li01,wyithe01,keeton01,takahashi01},
and (iii) observational confrontation for an individual object may
suffer from the specific selection function and the scatter from the
mean profile, and thus the statistical average over the cosmological
volume is important.

Several authors have already examined the effect of the inner profile of
dark halo on giant luminous arcs
\citep{wu93,miralda93a,miralda93b,hamana97,hattori97,williams99,
molikawa99,meneghetti01}.  In particular \citet{bartelmann98} suggested
a strong dependence of arc statistics on cosmological parameters because
of the different core structure of dark halos for different cosmological
model.  \citet{keeton01} found that the number of predicted lenses is
strongly correlated with core mass fraction, which results in a strong
degeneracy between the inner slope of the central cusp and the dark
matter concentration.  \citet{molikawa01} pointed out that the number
ratio of tangential and radial arcs produced by a given cluster is
tightly correlated with the inner slope; on the basis of eleven
tangential and three radial arcs for six clusters, they conclude that
the central cusp $\propto r^{-1.5}$ is indeed favored. Although the 
existing samples of clusters are somewhat heterogeneous and do not
satisfy well-defined selection criteria, this indicates that the arc
statistics become useful probes of the core structure of the dark halos.
Therefore we present a first systematic study of the effects of the dark
halo profiles on the gravitational tangential and radial arc statistics.
In particular we take proper account of the magnification bias, the
finite size of the source galaxies, and the luminosity distribution and
evolution of source galaxies.

The rest of the paper is organized as follows. Section \ref{sec:profile}
briefly summarizes the main properties of the generalized NFW profile,
and \S \ref{sec:arcstat} presents an analytic formalism of the number
count of arcs. Our predictions for the arc statistics are shown in \S
\ref{sec:results}. Finally we summarize the conclusions and discuss
further implications in \S \ref{sec:conclusions}.

\section{Description of the Density Profiles of Dark Matter  Halos}
\label{sec:profile}

\subsection{Generalized NFW Profile}
\label{subsec:NFW}
Throughout the paper, we adopt a generalized NFW profile for dark matter
halos of a form \citep{jing00a}:
\begin{equation}
\rho(r)=\frac{\rho_{\rm crit}\delta_{\rm c}}
{\left(r/r_{\rm s}\right)^\alpha\left(1+r/r_{\rm s}\right)^{3-\alpha}},
\label{nfw}
\end{equation}
where $r_{\rm s}$ is a scale radius and $\delta_{\rm c}$ is a
characteristic density. The profile with $\alpha=1$ corresponds to the one
NFW proposed, and that with $\alpha=1.5$ agree with the inner profile
claimed by \citet{moore99} and \citet{fukushige01}. The shape of halos
is also characterized by the concentration parameter, which is defined
as the ratio of the size of the halo to the scaled radius $r_{\rm s}$.
Originally NFW used $r_{200}$, the radius of halo where the mean inner
density reaches 200 times the critical density of the universe.  Rather
we follow \citet{bullock01} and adopt the definition:
\begin{equation}
c_{\rm vir}(M_{\rm vir})\equiv 
\frac{r_{\rm vir}(M_{\rm vir})}{r_{\rm s}(M_{\rm vir})}.
\label{concentration}
\end{equation}
The virial radius, $r_{\rm vir}$, in the above expression is defined
through the overdensity $\Delta_{\rm vir}$ at the virialization epoch
$z_{\rm vir}$ as
\begin{equation}
M_{\rm vir}=\frac{4\pi}{3}\Delta_{\rm vir}
    \bar{\rho}(z_{\rm vir})r_{\rm vir}^3 ,
\label{virial}
\end{equation}
where $\bar{\rho}(z_{\rm vir})$ denotes the mean density of the universe
at virialization. Once the density parameter $\Omega_0$ and the
cosmological constant $\lambda_0$ are specified, 
the value of $\Delta_{\rm vir}$ can be computed using
the nonlinear spherical collapse model. We use the following 
formulae:
\begin{eqnarray}
  \Delta_{\rm vir} = \left\{
      \begin{array}{ll}
        4\pi^2\displaystyle\frac{(\cosh \eta_{\rm vir}-1)^3}
{(\sinh \eta_{\rm vir}-\eta_{\rm vir})^2} &
        \mbox{($\Omega_0<1, \lambda_0=0$)}, \\ 
        18\pi^2(1+0.4093\omega_{\rm vir}^{0.9052}) &
        \mbox{($\Omega_0<1, \lambda_0=1-\Omega_0$)} , 
      \end{array}
   \right. 
\label{eq:deltavir}
\end{eqnarray}
where $\eta_{\rm vir}\equiv\cosh^{-1}(2/\Omega_{\rm vir}-1)$,
$\omega_{\rm vir}\equiv 1/\Omega_{\rm vir}-1$, and the density parameter
at virialization is 
\begin{equation}
\Omega_{\rm vir} = \frac{\Omega_0(1+z_{\rm vir})^3}
{\Omega_0(1+z_{\rm vir})^3+(1-\Omega_0-\lambda_0)(1+z_{\rm
vir})^2+\lambda_0} .
\end{equation}
The approximation to $\Delta_{\rm vir}$ in equation (\ref{eq:deltavir})
for $\lambda_0=1-\Omega_0$ is obtained by \citet{kitayama96}.

Equations (\ref{nfw}) and (\ref{virial}) imply that the characteristic
density $\delta_{\rm c}$ is related to the concentration parameter
$c_{\rm vir}$ as
\begin{equation}
\delta_{\rm c}=\frac{\Delta_{\rm vir}\Omega_{\rm vir}}{3}
\frac{c_{\rm vir}^3}{A(c_{\rm vir})},
\end{equation}
where $A(c_{\rm vir})$ is 
\begin{equation}
 A(c_{\rm vir})=\frac{c_{\rm vir}^{3-\alpha}}{3-\alpha}
\, {}_2F_1\left(3-\alpha, 3-\alpha; 4-\alpha; -c_{\rm vir}\right),
\end{equation}
with ${}_2F_1\left(a, b; c; x\right)$ being the hypergeometric function 
\citep[e.g.,][]{keeton01}.

\subsection{Concentration Parameter}
\label{subsec:concentration}

\citet{navarro97} and \citet{bullock01} have extensively examined the
cosmological model dependence and redshift evolution of the
concentration parameter from N-body simulations, adopting the profile
(\ref{nfw}) with $\alpha=1$. Since we consider models with $\alpha
\not=1$ as well, we have to generalize their results. For this purpose,
we follow \citet{keeton01}. They first define the radius $r_{-2}$ at
which the logarithmic slope of the density profile is $-2$, i.e.,
$d\ln\rho/d\ln r=-2$. For the profile of equation (\ref{nfw}), $r_{-2}=
(2-\alpha)r_{\rm s}$, and thus the corresponding concentration parameter
reduces to
\begin{equation}
c_{-2}\equiv \frac{r_{\rm vir}}{r_{-2}} = \frac{1}{2-\alpha}c_{\rm vir}.
\end{equation}
Then they point out the importance of the scatter of the concentration
parameter \citep{jing00b,bullock01} on the lensing statistics, and model
the probability distribution function of $c_{-2}$ as a log-normal
function:
\begin{equation}
 p(c_{-2})dc_{-2}=\frac{1}{\sqrt{2\pi}\sigma_c}
\exp\left[-\frac{(\ln c_{-2}-\ln c_{-2, {\rm median}})^2}
{2\sigma_c^2}\right]d\ln c_{-2},
\end{equation}
with $\sigma_c=0.18$. \citet{jing00b} reported that this dispersion is
fairly insensitive to the cosmological model parameters.  Finally we
introduce the scaling of $c_{-2,{\rm median}}$ according to the
simulations by \citet{bullock01}:
\begin{equation}
c_{-2, {\rm median}}=c_{\rm norm}\frac{1}{1+z}
\left(\frac{M_{\rm vir}}{10^{14}h^{-1}M_{\odot}}\right)^{-0.13} ,
\end{equation}
where $h$ denotes the Hubble constant in units of 100 km/s/Mpc.
\citet{bullock01} estimate $c_{\rm norm}\sim 8$ from their simulations with
$(\Omega_0, \lambda_0)=(0.3, 0.7)$.  In the statistical analyses
presented below, we parameterize the halo profiles by the amplitude
$c_{\rm norm}$ and the inner slope $\alpha$, assuming that the above
description is applicable equally well to the three cosmological models
that we consider.

\section{Tangential and Radial Arc Statistics}
\label{sec:arcstat}

\subsection{Lens Equations}
\label{subsec:basic}
We denote the image position in the lens plane by $\vec{\xi}$ and the
source position in the source plane by $\vec{\eta}$.  For the spherical
symmetric profile (\ref{nfw}), the lens equation
\citep[e.g.,][]{schneider92} reduces to
\begin{equation}
y=x-b\,f(x),
\end{equation}
where $x=|\vec{\xi}|/r_{\rm s}$, $y=|\vec{\eta}|D_{\rm OL}/r_{\rm
s}D_{\rm OS}$, and $D_{\rm OL}$ and $D_{\rm OS}$ denote the angular
diameter distances from the observer to the lens and the source planes,
respectively. The factors $b$ and $f(x)$ are related to the dark halo
profile as follows:
\begin{eqnarray}
b &=&\frac{4\rho_{\rm crit}\delta_{\rm c}r_{\rm s}}{\Sigma_{\rm crit}}, \\
f(x) &=& \frac{1}{x}\int_0^\infty dz \int_0^x dx' 
\frac{x'}{\left(\sqrt{x^{'2}+z^2}\right)^\alpha
\left(1+\sqrt{x^{'2}+z^2}\right)^{3-\alpha}},
\end{eqnarray}
where $\Sigma_{\rm crit}$ is the critical surface mass density:
\begin{equation}
\label{sigma_crit}
 \Sigma_{\rm crit}=
\frac{c^2}{4\pi G}\frac{D_{\rm OS}}{D_{\rm OL}D_{\rm LS}}.
\end{equation}

\subsection{Number of Arcs per Halo}
\label{subsec:perhalo}

We consider the distortion of images of source galaxies due to the
spherical halo lensing, neglecting the intrinsic ellipticity of those
galaxies.  In this case the tangential and radial stretching factors of a
source image at the source position $x$ with respect to the center
of the lens halo are simply given by $\mu_{\rm t}(x)\equiv(y/x)^{-1}$, and
$\mu_{\rm r}(x)\equiv(dy/dx)^{-1}$ because of the spherical symmetry of
halos. In terms of these, we define the tangential and radial arcs as
those satisfying
\begin{eqnarray}
T(x)&=&\left|\frac{\mu_{\rm t}(x)}{\mu_{\rm r}(x)}\right|
 >\epsilon_{\rm th}\;\;\;{\rm(tangential~arc),}
\label{tth}\\
R(x)&=&\left|\frac{\mu_{\rm r}(x)}{\mu_{\rm t}(x)}\right|
>\epsilon_{\rm th}\;\;\;{\rm(radial~arc).}
\label{rth}
\end{eqnarray}
We adopt $\epsilon_{\rm th}=4$ for the threshold of the length-to-width 
ratio. This value is different from the canonical threshold for giant 
luminous arcs, $\epsilon_{\rm th}=10$ adopted by \citet{wu93}, because 
most radial arcs presently observed have a rather small length-to-width ratio. 

The projected areas around a given halo of mass $M$ at $z_{\rm L}$
 satisfying equations (\ref{tth}) and (\ref{rth}) yield the
 cross sections $\sigma(M, z_{\rm L}, z_{\rm S})$ for tangential and
 radial arcs of circular galaxies located at $z_{\rm S}$, respectively.
 Actually the above definition ignores the finite size of the source
 galaxies. If the size of the area on the source plane, $\Delta\eta$,
 satisfying the conditions (\ref{tth}) or (\ref{rth}) is smaller than
 that of source, however, observable arcs are not produced (e.g.,
 Schneider et al. 1992). Since the smallest galaxy size that is observed
 as an arc (Hattori et al. 1997) roughly corresponds to $\eta_{\rm
 crit}=1h^{-1}$ kpc, we set $\sigma(M, z_{\rm L}, z_{\rm S})=0$ for
 $\Delta\eta<\eta_{\rm crit}$.

\begin{figure}
\plotone{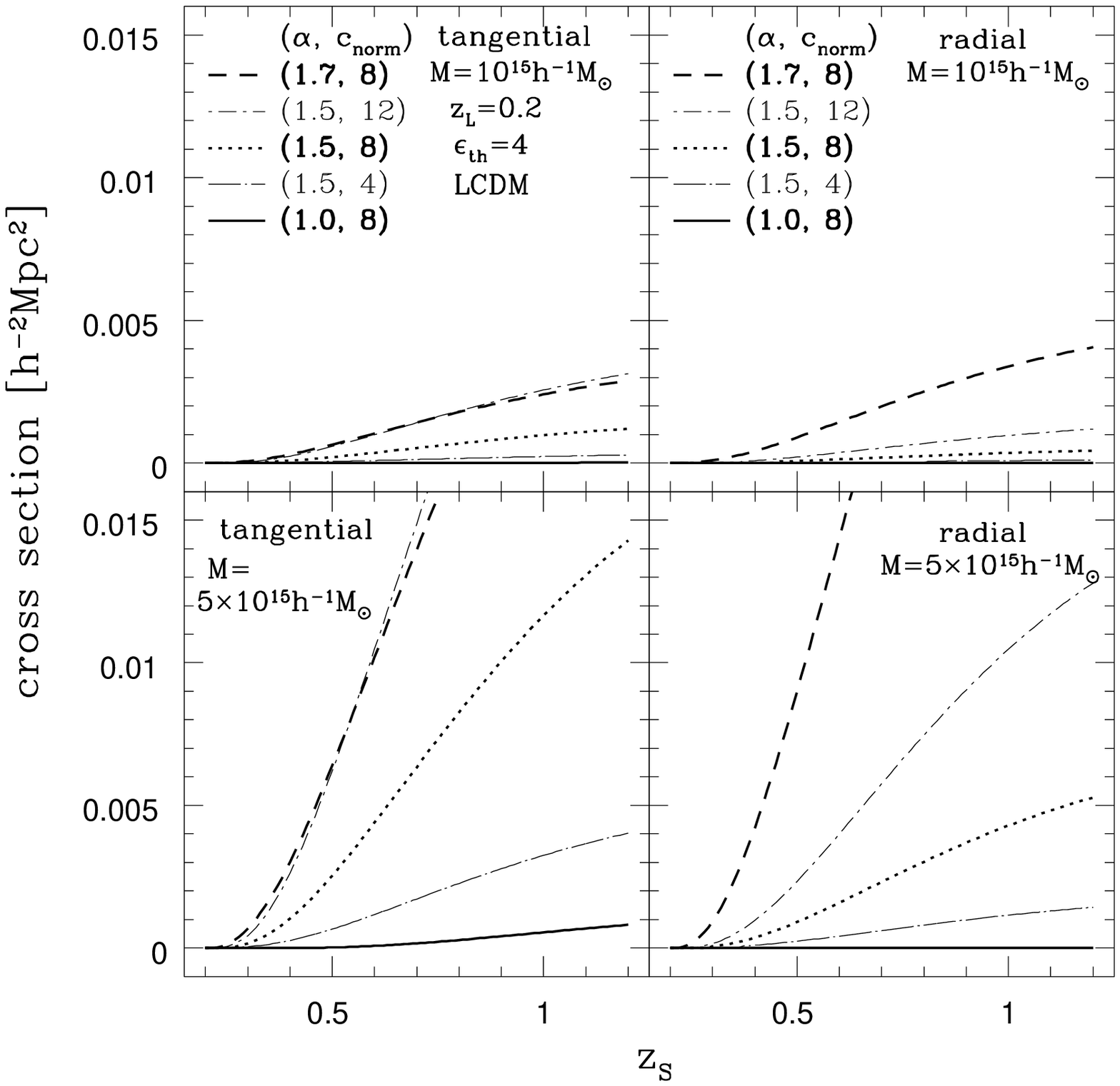} \caption{The lensing cross sections for tangential
({\it left panels}) and radial ({\it right panels}) arcs against the
source redshift $z_{\rm S}$ for an $\Omega_0$=0.3 and $\lambda_0$=0.7
model.  The redshift of the lensing halo is fixed as $z_{\rm L}=0.2$.
Different lines correspond to the different sets of $\alpha$ and $c_{\rm
norm}$.  Top and bottom panels indicate the results for
the  halo mass of $10^{15}h^{-1}M_\odot$ and 
$5\times 10^{15}h^{-1}M_\odot$, respectively.} \label{fig:cross}
\end{figure}

Figure \ref{fig:cross} shows the lensing cross sections for tangential
and radial arcs as a function of source redshift $z_{\rm S}$. We present
the cases that the mass of the lens halo is $M=10^{15}h^{-1}M_{\odot}$
and $M=5\times 10^{15}h^{-1}M_{\odot}$, because radial arcs are not
formed for $M\sim10^{14}h^{-1}M_{\odot}$ unless $\alpha$ or $c_{\rm
norm}$ is unrealistically large.  The redshift of the lens halo is fixed
to $z_{\rm L}=0.2$. These plots show that the lensing cross section
significantly increases as $\alpha$ and/or $c_{\rm norm}$ become larger.
Furthermore the cross sections for tangential and radial arcs depend on
these two parameters rather differently.  In turn, separate consideration
of tangential and radial arcs yields useful constraints on the core
structure of dark halos.

Once the relevant cross section is computed, one can calculate the number
of arcs produced by a halo of mass $M$ at redshift $z_{\rm L}$:
\begin{equation}
N(M, z_{\rm L})=\int^{z_{\rm S, max}}_{z_{\rm L}}dz_{\rm S}\,\sigma(M, z_{\rm L}, z_{\rm S})\frac{c\,dt}{dz_{\rm S}}(1+z_{\rm S})^3\int_{L_{\rm min}}^\infty dL\,n_{\rm g}(L, z_{\rm S}),
\label{perhalo}
\end{equation}
where $\sigma(M,z_{\rm L},z_{\rm S})$ is the tangential or radial cross 
section in the source plane, $c\,dt/dz_{\rm S}$ denotes the proper
differential distance at $z_{\rm S}$:
\begin{equation}
\frac{c\,dt}{dz_{\rm S}}=\frac{c}{H_0}\frac{1}{(1+z_{\rm S})\sqrt{\Omega_0(1+z_{\rm S})^3+(1-\Omega_0-\lambda_0)(1+z_{\rm S})^2+\lambda_0}},
\end{equation}
and $n_{\rm g}(L,z_{\rm S})$ denotes the luminosity function of source 
galaxies. 

We incorporate the redshift evolution of the luminosity function of
galaxies for $z\lesssim1$ adopting the empirical fit by
\citet{broadhurst88}:
\begin{equation}
\log\,\phi (L, z)=\log\phi (L, 0)+(0.1z+0.2z^2)
\log\left[\frac{\phi(L, 0)}{\phi(L_{\rm max}, 0)}\right],
\label{lum}
\end{equation}
where $L_{\rm max}$ is the bright-end luminosity corresponding to
$M_{\rm max}=-22.0+5\log h$. Therefore we have to restrict our
consideration for source galaxies up to $z_{\rm S,max}=1$.

For the local luminosity function $\phi(L, 0)$, we use the Schechter
function normalized to the two degree field (2dF) galaxy redshift
survey:
\begin{equation}
\phi(L, 0)dL=\phi^*\left(\frac{L}{L^*}\right)^\alpha\exp\left(-\frac{L}{L^*}\right)\frac{dL}{L^*}
\end{equation}
where $\phi^*=0.0169h^{-3} {\rm Mpc^3}$, $\alpha=-1.28$, and
$M^*=-19.73+5\log h$ \citep{folkes99}.  While those values are derived
assuming $\Omega_0=1$, we compute the values in different cosmological
models by applying an appropriate scaling so that the observed galaxy
number counts versus their flux is unchanged.

To evaluate equation (\ref{perhalo}), we also need the lower bound of
the luminosity, $L_{\rm min}$, which depends on the magnification of
arcs and the limiting magnitude of the sample. First, the magnification of
arcs becomes
\begin{equation}
\mu(x)=\left|\mu_{\rm t}(x)\mu_{\rm r}(x)\right|=T(x)\left\{\mu_{\rm r}(x)\right\}^2=R(x)\left\{\mu_{\rm t}(x)\right\}^2,
\end{equation}
where $x$ denotes the position of arcs in the lens plane.  The quantity
$\mu$ formally diverges in the case of the point source. In practice,
however, $\mu(x)$ saturates at about the value which corresponds to the
value of the position deviating from the critical curve by the source
size (e.g., Schneider et al. 1992). Therefore we assume that all arcs
are magnified by the factor of $\epsilon_{\rm th}\left\{\mu_{\rm r}(x_{\rm
t})\right\}^2$ for tangential arc and $\epsilon_{\rm th}\left\{\mu_{\rm
t}(x_{\rm r})\right\}^2$ for radial arcs, where $x_{\rm t}$ and $x_{\rm
r}$ is the positions of tangential and radial critical curves. Second, we
use $m_{\rm B}<26.5$ as the B-band limiting magnitude of the arc
observation. The apparent magnitude can be translated to the luminosity
if we employ the K-correction in B-band:
\begin{equation}
K(z)=-0.05+2.35z+2.55z^2-4.89z^3+1.85z^4
\end{equation}
for spiral galaxies \citep{king85}. Taking both effects into account, 
$L_{\rm min}$ becomes
\begin{equation}
\frac{L_{\rm min}}{L^*}=\frac{10^{-0.4(m_{\rm B,max}-m^*)}}{\epsilon_{\rm th}\left\{\mu_{\rm r}(x_{\rm t})\right\}^2},
\end{equation}
\begin{equation}
m^*=M^*+5\log\left(\frac{D_{\rm lum}(z_{\rm s})}
{10{\rm pc}}\right)+K(z_{\rm S}),
\end{equation}
in the case of tangential arcs, where $D_{\rm lum}$ is the luminosity distance.

\subsection{Total Number of Arcs}
\label{subsec:total}

Finally we calculate the total number of arcs by integrating equation
 (\ref{perhalo}) as:
\begin{equation}
N_{\rm tot}=\int_{z_{\rm L,min}}^{z_{\rm L, max}}dz_{\rm L}
\int_{M_{\rm min}(z_{\rm L})}^\infty dM\,N(M, z_{\rm L})
\, n_{\rm PS}(M, z_{\rm L}) (1+z_{\rm L})^3 4\pi D_{\rm OL}^2\frac{c\,dt}{dz_{\rm L}},
\label{eq:ntot}
\end{equation}
where $n_{\rm PS}$ is the comoving number density of halos. We use
Press-Schechter mass function \citep{press74}:
\begin{equation}
n_{\rm PS}(M, z)=\sqrt{\frac{2}{\pi}}\frac{\bar{\rho}(z=0)}{M}\frac{\delta_0(z)}{\sigma_M^2}\left|\frac{d\sigma_M}{dM}\right|\exp\left[-\frac{\delta_0^2(z)}{2\sigma_M^2}\right],
\label{ps}
\end{equation}
where $\sigma_M$ is the rms of linear density fluctuation on mass scale
$M$ at $z=0$ and $\delta_0(z)$ is the critical linear density contrast 
given by
\begin{equation}
\delta_0(z)=\frac{3}{20}\frac{\left(12\pi\right)^{2/3}}{D(z)},
\end{equation}
with $D(z)$ being the linear growth rate normalized to unity at $z=0$.
We consider two selection functions, i.e., the minimum mass of integration
$M_{\rm min}(z_{\rm L})$ for definiteness; the first adopts the
constant minimum mass independent of $z_{\rm L}$, and the other
corresponds to the X-ray survey with the surface brightness flux limit
of $S_{\rm min}$. Throughout the paper, we use $0.5-2.0$ keV band for
the flux. Assuming the conventional one-to-one correspondence between
dark halos as X-ray clusters, one can relate $M_{\rm min}(z_{\rm L})$
with $S_{\rm min}$ and we use the relation shown in \citet{suto00}.

\section{Results}
\label{sec:results}

\subsection{Number Counts of Arcs}
\label{subsec:count}

In the specific examples presented below, we consider three
representative cosmological models dominated by cold dark matter (CDM);
Lambda CDM (LCDM), Standard CDM (SCDM), and Open CDM (OCDM) with
$(\Omega_0,~\lambda_0,~h,~\sigma_8) = (0.3,~0.7,~0.7,~1.04)$, $(1.0,~0.0,~
0.5,~0.56)$, and $(0.45,~0.0,~0.7,~0.83)$, respectively. The amplitude
of the mass fluctuation, $\sigma_8$, smoothed over the top-hat radius of
$8h^{-1}$ Mpc, is normalized so as to reproduce the X-ray luminosity and
temperature functions of clusters \citep{kitayama97}.

\begin{figure}
\plotone{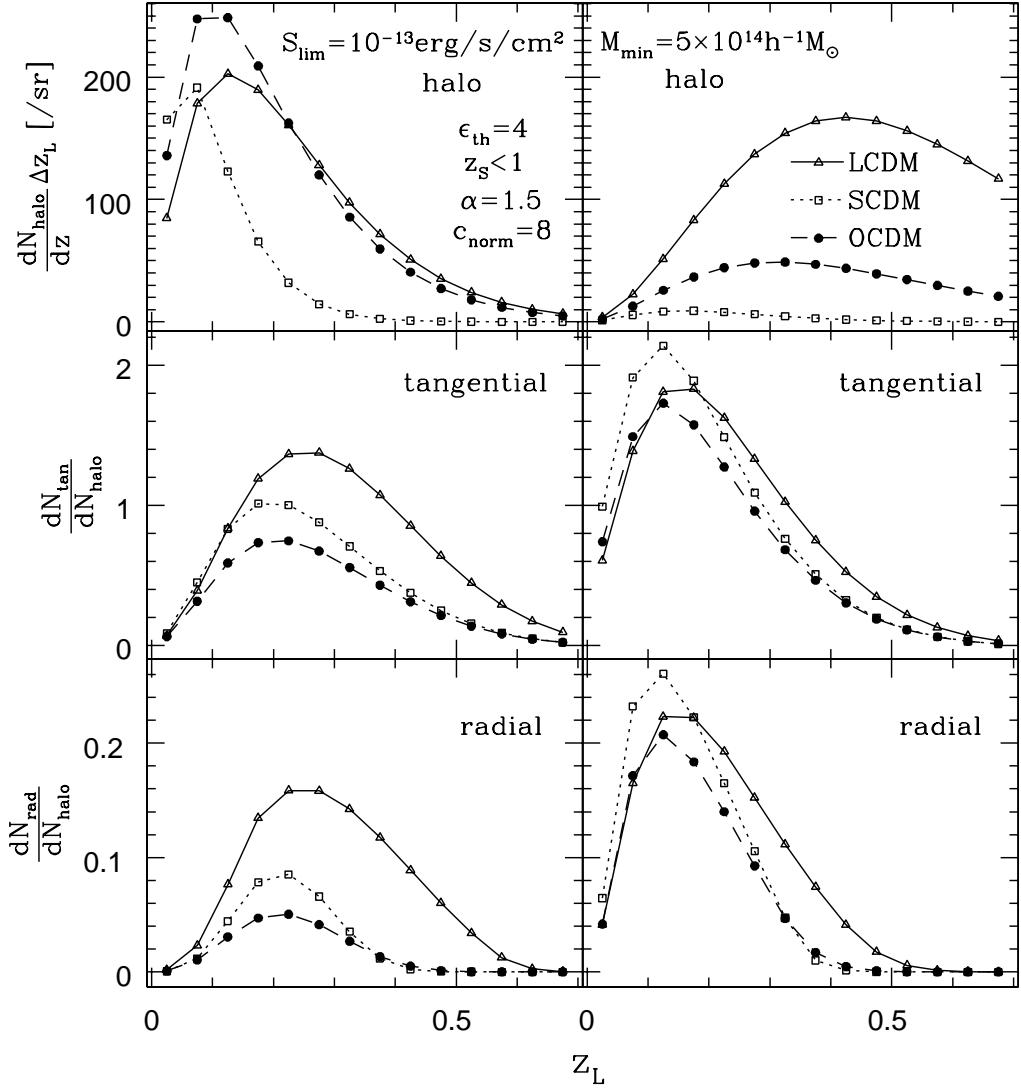} \vspace{-10mm}
\caption{Redshift distribution of the numbers of halos
per steradian ({\it top panels}), tangential arcs ({\it middle panels}),
and radial arcs ({\it bottom panels}).  The numbers of tangential and
radial arcs in the middle and bottom panels are divided by the number of
halos plotted in the top panels.  We adopt $\alpha=1.5$, $c_{\rm norm}=8$,
and $z_{\rm S}<1$, and plot those numbers between $z_{\rm L} - \Delta
z_{\rm L}/2$ and $z_{\rm L} + \Delta z_{\rm L}/2$ with $\Delta z_{\rm
L}=0.05$.  Open triangles, open squares, and filled circles indicate the
results for LCDM, SCDM, and OCDM.  The left and right panels correspond
to the X-ray flux-limited ($S_{\rm lim}=10^{-13}{\rm erg/s/cm^2}$) and
the mass-limited ($M_{\rm min}=5\times10^{14}h^{-1}M_\odot$) samples.}
\label{fig:nz}
\end{figure}

Figure \ref{fig:nz} plots the number of halos per steradian ({\it Top
panels}), tangential arcs ({\it Middle panels}), and radial arcs ({\it
Bottom panels}) between $z_{\rm L} - \Delta z_{\rm L}/2$ and $z_{\rm L} +
\Delta z_{\rm L}/2$ with $\Delta z_{\rm L}=0.05$, for $\alpha=1.5$, $c_{\rm
norm}=8$, and $z_{\rm S}<1$. The triangles, open squares, and filled
circles indicate the results for LCDM, SCDM and OCDM.  The numbers of
arcs in the middle and bottom panels are divided by the number of halos
plotted in the top panels.  The left and right panels correspond to the
X-ray flux-limited ($S_{\rm lim}=10^{-13}{\rm erg/s/cm^2}$) and the
mass-limited ($M_{\rm min}=5\times10^{14}h^{-1}M_\odot$) samples.  For a
given mass and profile of a halo, these plots indicate the range of
$z_{\rm L}$ which mostly contributes to the formation of arcs.  Both
tangential and radial arcs are efficiently formed around $z_{\rm L}\sim 0.2$
and $z_{\rm L}\sim 0.1$ for flux-limited and mass-limited samples,
respectively.

\begin{figure}
\plotone{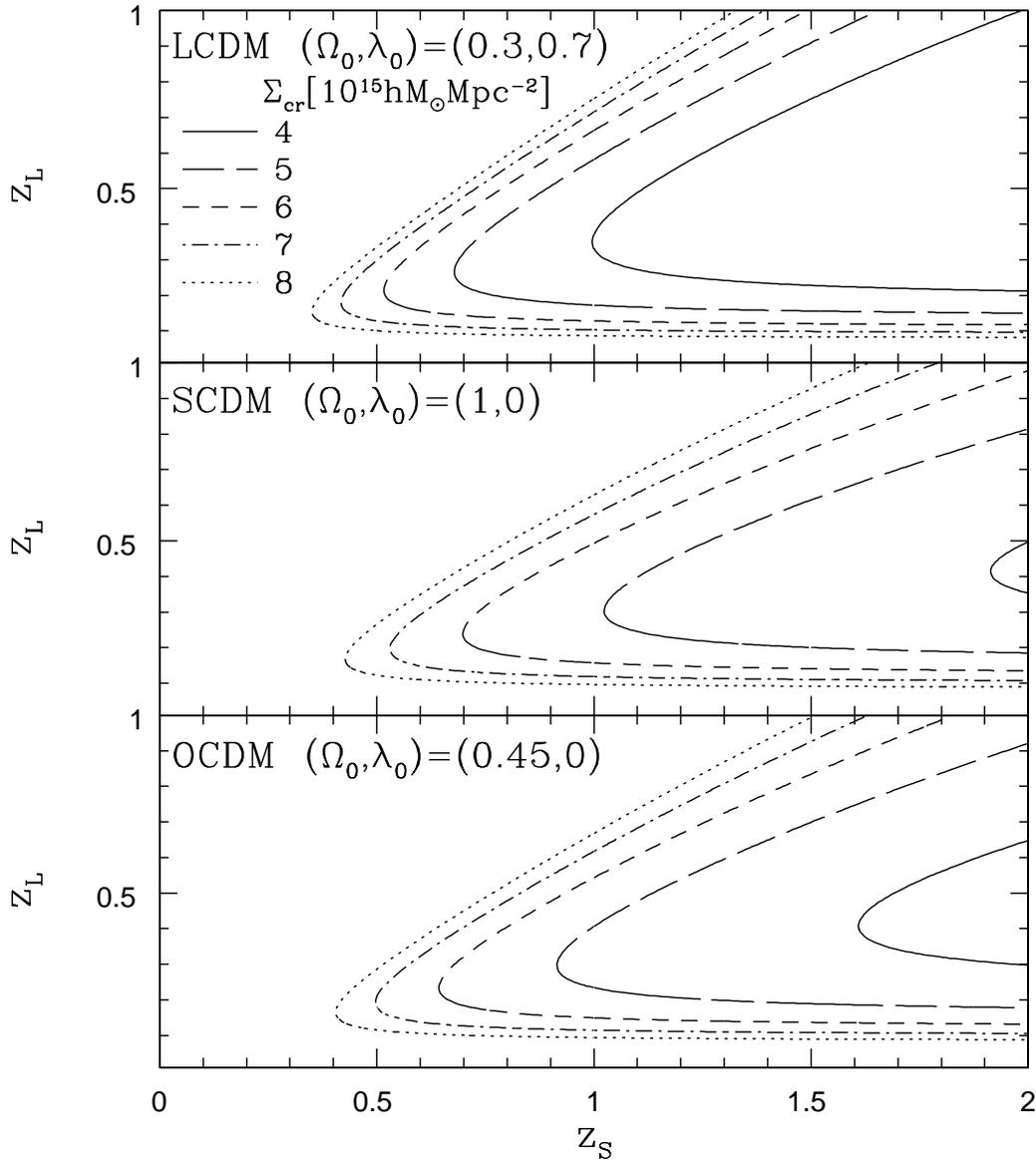} \caption{Contour of the critical surface density
$\Sigma_{\rm crit}$ on the $z_{\rm S} - z_{\rm L}$ plane for three
cosmological models; LCDM ({\it top panel}), SCDM ({\it middle
panel}), and OCDM ({\it bottom panel}). The 
contour levels for each line are
 $\Sigma_{\rm crit}=4$ ({\it solid}), $5$ ({\it long-dashed}),
$6$ ({\it short-dashed}), $7$ ({\it dash-dotted}) and $8$ ({\it
dotted}), in units of $10^{15}hM_\odot{\rm Mpc^{-2}}$.}
\label{fig:sigma_crit}
\end{figure}

Note the strong dependence on cosmological parameters. The number of
halos per steradian as a function of $z_{\rm L}$ depends on both the
volume of the universe up to $z_{\rm L}$ and the evolution of the mass
function.  The former mainly explains why the halos are most abundant in
LCDM at higher redshifts, while the latter accounts for earlier
declining of halo numbers in SCDM (top panels in Fig. \ref{fig:nz}).
Although this behavior is already well-known in the study of cluster
abundance \citep{kitayama98}, we emphasize that the number of arcs {\it per
halo} also increases with the presence of the cosmological constant
(middle and bottom panels in Fig. \ref{fig:nz}), as is pointed out by
\citet{wu96}. This directly comes from the dependence of the critical
surface mass density (eq. [\ref{sigma_crit}]). As plotted in 
Figure \ref{fig:sigma_crit}, the value of $\Sigma_{\rm crit}$ is
smallest in LCDM for given $z_{\rm S}$ and $z_{\rm L}$, i.e., the
lensing probability is largest for a given halo profile. At $z\lesssim0.1$,
SCDM produces more abundant arcs than LCDM because of the higher density of
halos for the same halo mass, i.e., $\Delta_{\rm vir}\bar{\rho}(z_{\rm
vir})$ becomes the largest in SCDM cosmology (see eq. [\ref{eq:deltavir}]).

\begin{figure}
\plotone{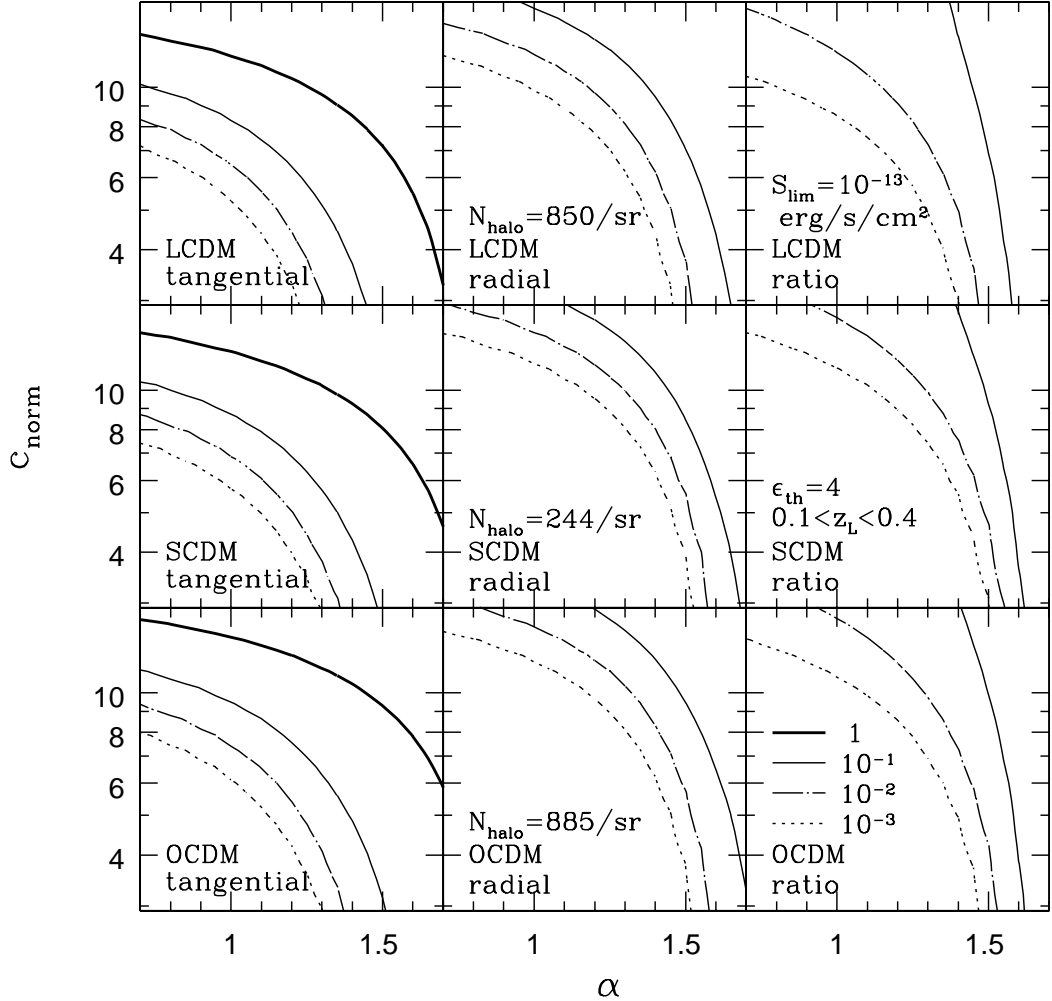} \caption{Predicted numbers per halo for tangential
({\it left panels}) and radial ({\it center panels}) arcs per halos, and
their ratio $W\equiv N_{\rm tot, rad}/N_{\rm tot, tan}$ ({\it right
panels}) on $\alpha$ and $c_{\rm norm}$ plane.  We consider X-ray
flux-limited samples with $S_{\rm lim}=10^{-13} {\rm erg/s/cm^2}$ in
three representative cosmological models; LCDM ({\it top panels}), SCDM
({\it middle panels}), and OCDM ({\it bottom panels}). The number of
halo $N_{\rm halo}$ is also shown for reference. The levels of contour
are $1$ ({\it thick solid}), $10^{-1}$ ({\it thin solid}), $10^{-2}$
({\it dash-dotted}) and $10^{-3}$ ({\it dotted}).}  \label{fig:sx13}
\end{figure}

Turn next to the dependence of the halo profiles ($c_{\rm norm}$ and
$\alpha$) on the arc statistics.  Figure \ref{fig:sx13} displays the
contour of arc statistics for the flux-limited sample with $S_{\rm
lim}=10^{-13} {\rm erg/s/cm^2}$; LCDM ({\it Top panels}), SCDM ({\it
Middle panels}), and OCDM ({\it Bottom panels}).  The left and center
panels indicate the number of tangential and radial arcs {\it per
halos}, and the right panels plot the number ratio of radial to
tangential arcs.  Here we integrate equation (\ref{eq:ntot}) over the
lens redshift of $0.1<z_{\rm L}<0.4$ when arcs are efficiently formed as
we discussed above.  Clearly both $\alpha$ and $c_{\rm norm}$
significantly influence the arc statistics, an order of magnitude more
than the cosmological parameters.

If we use the same statistics:
\begin{equation}
 W\equiv \frac{N_{\rm tot, rad}}{N_{\rm tot, tan}}
\label{w}
\end{equation}
proposed by \citet{molikawa01} originally for a single cluster, we
confirm that the cosmological model dependence is extremely weak even
after the statistical average over the redshift.  In particular, the
number ratio $W$ is not so sensitive to $c_{\rm norm}$ and basically a
powerful indicator of the inner slope of the dark halo profile.  The
number of tangential arcs per halos, on the other hand, is more
sensitive to $c_{\rm norm}$ ({\it Left panels}). Thus combining both the
tangential and radial arc statistics, we can constrain both $\alpha$ and
$c_{\rm norm}$ simultaneously.

\begin{figure}
\plotone{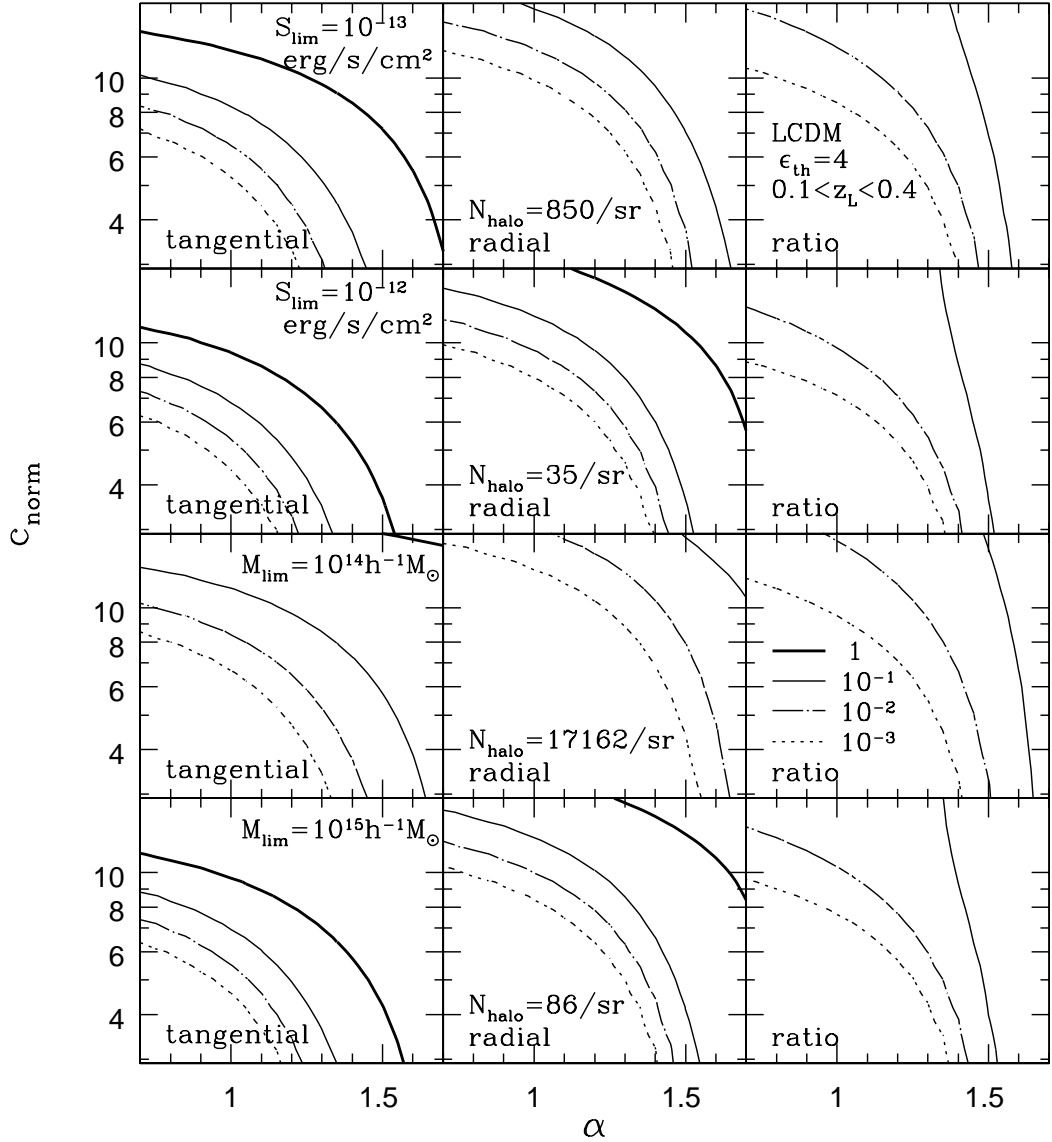} \caption{Effect of sample selection functions on the
arc statistics in the LCDM model. From top to bottom panels, X-ray
flux-limited samples with $S_{\rm lim}=10^{-13} {\rm erg/s/cm^2}$ and
$S_{\rm lim}=10^{-12} {\rm erg/s/cm^2}$, and mass-limited samples with
$M_{\rm min}=10^{14}h^{-1}M_\odot$ and $M_{\rm
min}=10^{15}h^{-1}M_\odot$.  The contour levels are the same as Figure
\ref{fig:sx13}.}  \label{fig:sm}
\end{figure}

\begin{figure}
\plotone{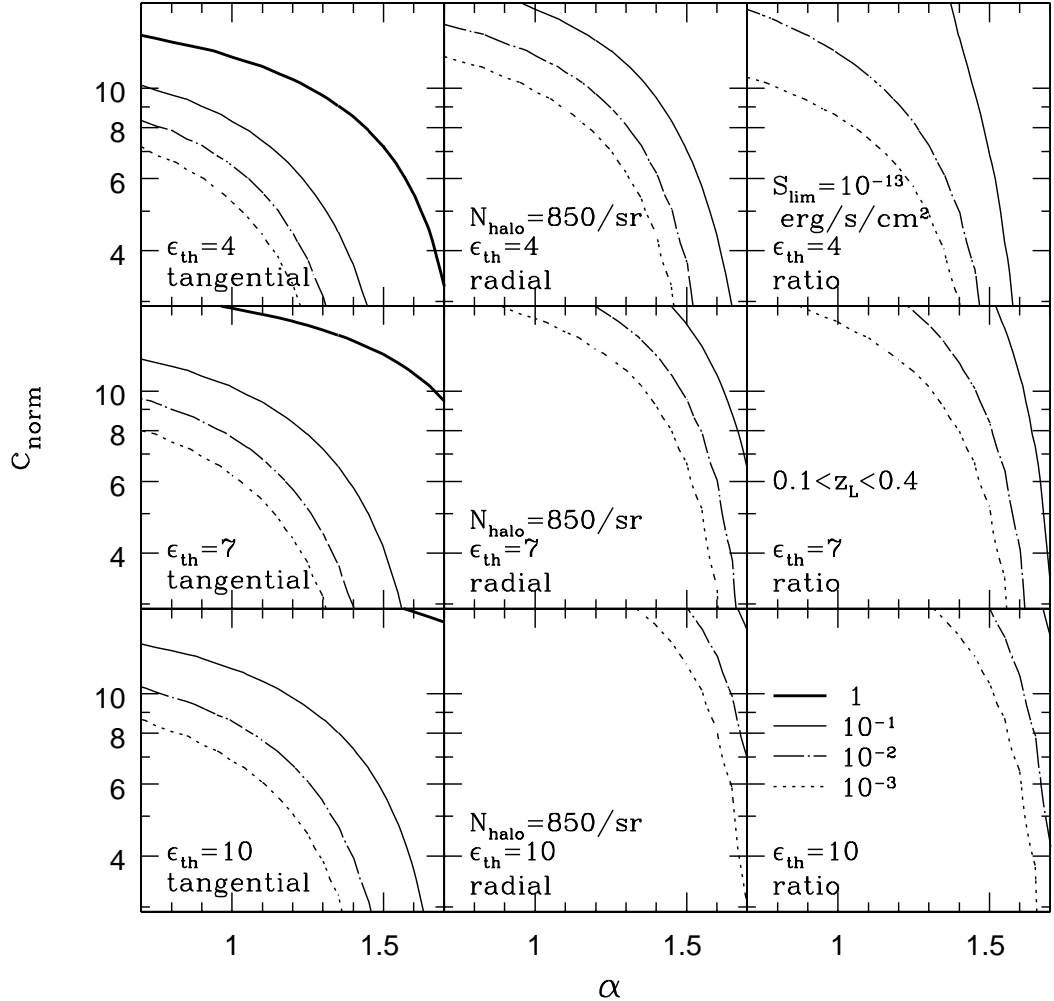} \caption{Effect of the different threshold of the
length-to-width ratio on the arc statistics for X-ray flux-limited
samples with $S_{\rm lim}=10^{-13} {\rm erg/s/cm^2}$ in the LCDM model;
$\epsilon_{\rm th}=4$ ({\it top panels}), $\epsilon_{\rm th}=7$ ({\it
middle panels}), and $\epsilon_{\rm th}=10$ ({\it bottom panels}). 
 The contour levels are the same as Figure
\ref{fig:sx13}.}  \label{fig:th}
\end{figure}

Next consider the effects of changing the selection criterion of both
halos and sources. Figure \ref{fig:sm} shows how the selection criterion
of lensing halos alters the prediction of arc statistics in the case
of LCDM; from top to bottom, $S_{\rm lim}=10^{-13} {\rm erg/s/cm^2}$,
$S_{\rm lim}=10^{-12} {\rm erg/s/cm^2}$, $M_{\rm
min}=10^{14}h^{-1}M_\odot$, and $M_{\rm min}=10^{15}h^{-1}M_\odot$.
While the different selection criterion yields a factor of 500
difference in the number of halos, the arc statistics {\it per halo} are
fairly robust, and remain the powerful discriminator of the halo
profile. Figure \ref{fig:th} displays the difference of the prediction by 
altering the threshold of the length-to-width ratio, for X-ray
flux-limited samples, $S_{\rm lim}=10^{-13} {\rm erg/s/cm^2}$, in LCDM
cosmology. As shown in these plots, changing the threshold mainly
changes the number of radial arcs and consequently changes the number
ratio $W$ drastically. This strong threshold dependence originates from
the finite source size effect and is not incorporated properly in 
\citet{molikawa01}. Therefore it is clear that the finite size of source
galaxies must be considered even in the case of the number ratio $W$.

\subsection{Uncertainties of the Predictions}
\label{subsec:unc}

The results presented above are based on a variety of model assumptions,
and we would like to examine the extent to which they affect the
conclusions. More specifically, we focus on the mass function for dark
halos, the size of source galaxies, and the evolution of the luminosity
function of source galaxies.

\begin{figure}
\plotone{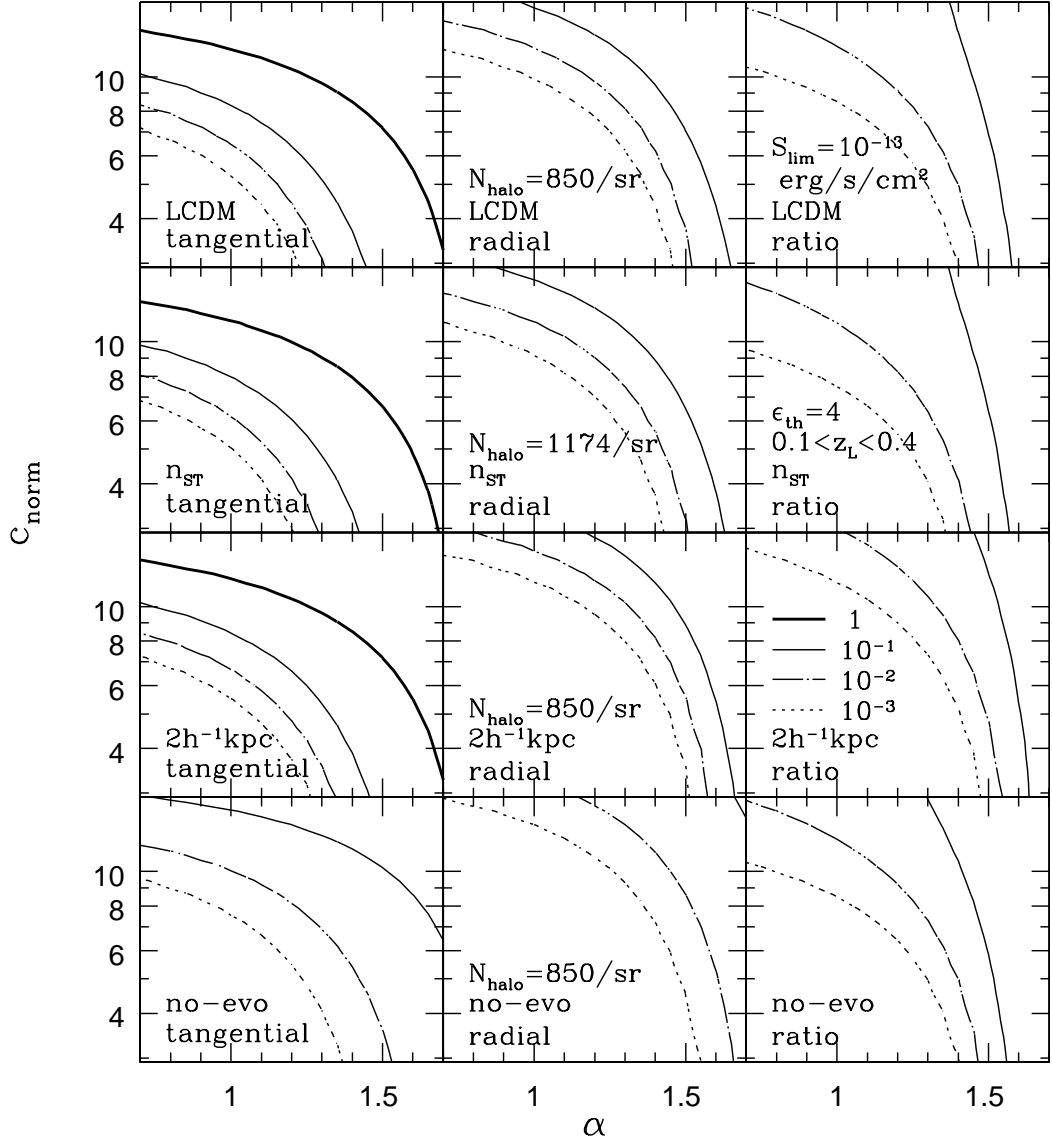} \caption{Sensitivity of our predictions on the adopted
model parameters for X-ray flux-limited samples with $S_{\rm
lim}=10^{-13} {\rm erg/s/cm^2}$ in the LCDM model.  {\it First row}
displays our fiducial model (same as Figure \ref{fig:sx13}).  {\it
Second row} adopts $n_{\rm ST}$ (eq. [\ref{st}]) instead of $n_{\rm PS}$
(eq. [\ref{ps}]) for the mass function of halos. {\it Third row} adopts
$\eta_{\rm crit}=2h^{-1}$ kpc instead of $\eta_{\rm crit}=1h^{-1}$ kpc
for the cutoff size of source galaxies. {\it Fourth row} assumes no
evolution in the luminosity function.}  \label{fig:unc}
\end{figure}

While the Press-Schechter mass function is widely used in various
cosmological predictions, recent numerical simulations
\citep[e.g.,][]{jenkins01} suggest that it underpredicts the massive
halos while overpredicts the less massive halos.  \citet{sheth99}
proposes an empirical correction for the effect as
\begin{equation}
n_{\rm ST}(M, z)=A\left[1+\left(\frac{\sigma_M}
{\sqrt{a}\delta_0(z)}\right)^{2p}\right]\sqrt{\frac{2a}{\pi}}
\frac{\bar{\rho}(z=0)}{M}\frac{\delta_0(z)}{\sigma_M^2}\left|
\frac{d\sigma_M}{dM}\right
|\exp\left[-\frac{a\delta_0^2(z)}{2\sigma_M^2}\right],
\label{st}
\end{equation}
where $a=0.707$, $p=0.3$, and $A=0.322$.  Also we consider the cases of
the twice larger threshold for $\eta_{\rm crit}=2h^{-1}$ kpc and
no-evolution luminosity function, i.e., $\phi(L, z)=\phi(L, 0)$.  The
results separately employing the above change are plotted in Figure
\ref{fig:unc}. Among them only the non-evolution model rather changes
the total number of arcs, but this may be too extreme. More importantly,
the ratio of the tangential and radial arcs, $W$, still remains
unchanged in practice even if the luminosity evolution is neglected.
The twice larger threshold model mainly changes the number of radial
arcs, and as a result this model also changes the number ratio $W$.
Thus we conclude that the arc statistics that we presented above 
are not so affected by the uncertainties of the models, and are fairly
robust discriminator of the halo profiles if the finite size of source
galaxies is correctly taken into account.

\subsection{Tentative Comparison with Observations}
\label{subsec:obs}
While there is no homogeneous sample for the arc survey available yet
that satisfies our selection criteria, it is tempting to make a
comparison with the existing data. \citet{luppino99} present
the results of imaging survey for gravitational lensing in a sample of
38 X-ray selected clusters of galaxies. From these clusters we choose a
subsample of 13 clusters which satisfy the condition $0.1<z<0.4$ and $S(
0.5-2.0 {\rm keV})>S_{\rm
lim}=10^{-12} {\rm erg/s/cm^2}$. Within these 13 clusters, 15 tangential
and 2 radial arcs with $\epsilon_{\rm th}=4$ are reported. We attempt to
draw cosmological implications by comparing these observational values
with our theoretical predictions.

\begin{figure}
\plotone{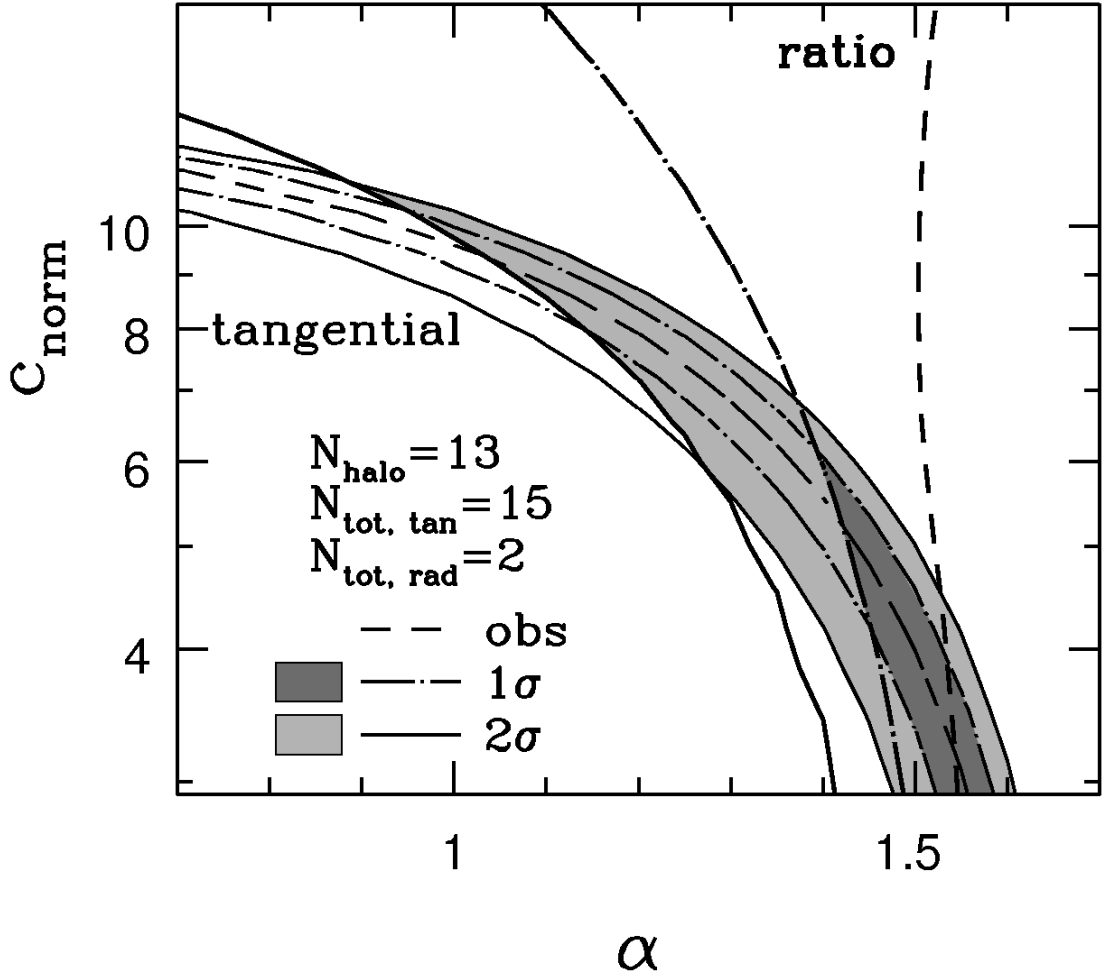} \caption{Tentative constraints on $\alpha$ and $c_{\rm
norm}$ from a sample of 13 clusters with $0.1<z<0.4$ and $S_{\rm
lim}=10^{-12} {\rm erg/s/cm^{-2}}$.  The LCDM model is assumed.  Dashed
lines represent the relation of $\alpha$ and $c_{\rm norm}$ which
reproduce the observed number for tangential arcs and the number ratio
 of radial to tangential arcs.  Dark shaded and light shaded regions
indicate the allowed regions combined from the two constraints taking
 account of the $1\sigma$ and $2\sigma$ statistical errors,
 respectively.}  \label{fig:obs}
\end{figure}

Figure \ref{fig:obs} shows our tentative comparison with observations,
assuming the LCDM cosmology and neglecting possible systematic errors
(e.g., intrinsic ellipticies of source galaxies, non-sphericities in
lensing halos, etc.). This result implies that dark matter halos should
have steep inner profiles as those predicted by simulations, $\alpha\sim
1.5$, but rather smaller concentration, $c_{\rm norm}\sim 4$.  While it
is premature to draw definite conclusions from the present comparison,
this analysis clearly exhibits the extent to which our current
methodology puts useful constraints on $\alpha$ and $c_{\rm norm}$
separately. Note again that the number of tangential arcs has a strong
degeneracy between $\alpha$ and $c_{\rm norm}$. Thus it is important to
combine the number of tangential arcs with the number ratio $W$ which is
mainly sensitive to $\alpha$. 
Although the present example does not show the strong constraint on the 
concentration parameter $c_{\rm norm}$, a more severe constraint within 
$\sim10\%$ accuracy at $1\sigma$ level will be  
obtained if one uses the cluster samples of $N_{\rm halo}\sim100$. This
would be achieved irrespective of the cosmological models.  
\section{Discussion and Conclusions}
\label{sec:conclusions}

In this paper, we study the arc statistics of gravitational lensing
 produced by dark matter halos in order to probe their density profile.
 Adopting the generalized NFW profile (\ref{nfw}), we describe a statistical
 method to predict the numbers of tangential and radial arcs as a
 function of the inner slope $\alpha$ and the concentration parameter
 $c_{\rm norm}$.  We incorporate several realistic effects including the
 magnification bias, the finite size of the source galaxies, and the
 luminosity distribution and evolution of source galaxies.  We find that
 the numbers of arcs sensitively depend on the values of $\alpha$ and
 $c_{\rm norm}$. In addition, the numbers of arcs, if divided by the
 corresponding number of halos, are almost insensitive to the underlying
 cosmological parameters. Therefore they prove to be a powerful
 discriminator of a family of halo density profiles suggested by recent
 numerical simulations.

\citet{molikawa01} proposed to use the number ratio of tangential and
radial arcs $W$ to probe the density profile. We confirm that the
ratio remains a useful {\it statistical} measure even after taking
account of the average over the cosmological mass function, the redshift
evolution, mass-dependence and the probability distribution of the
concentration parameter. We also show that the effects of
finite source size is important even in the case of the number ratio
$W$. On the other hand, $W$ is mainly sensitive to
the inner slope $\alpha$, and we argue that the complementary
information on $c_{\rm norm}$ can be obtained by combining the number of
tangential arcs per halo.

Our major conclusion that the cosmological model dependence of arc
statistics is much weaker than the profile parameters seems inconsistent
with the claim by \citet{bartelmann98}.  We note, however, that they use
a different value of concentration parameter for different cosmological
models. Thus we suspect that the claimed cosmological model dependence
actually reflects the sensitivity to the concentration parameter that we
discussed at length.

The preliminary comparison with observations suggests that dark matter
halos should have steep inner profiles. This comparison, however, is
inconsistent with the mass profile of CL0024-1654 reconstructed from
gravitational lensing images \citep{tyson98} and with the rotation
curves of low-surface brightness galaxies which indicate a flat core
\citep{salucci00,deBlok01}. Therefore it is still premature to draw any
strong conclusions, e.g., self-interacting dark matter model
\citep{spergel00}, at this point, and it is important to put many
constraints on the halo density profile from separate and independent 
analyses.
 
Definitely we have to improve the present methodology by taking account
of more realistic effects. Firstly, \citet{bartelmann95a} pointed out
that the number of arcs becomes significantly larger if the intrinsic
ellipticity distribution of source galaxies is taken into account. 
This will increase both the numbers of tangential and radial arcs. Thus
these quantitative estimate is important. Secondly, we neglect effects
of the galaxies in clusters. In particular, the central cD galaxies move
the radial arc closer to the center \citep{miralda95}, and will affect
especially the number of radial arcs. On the other hand, the effect of 
cluster galaxies seems to be small and enhances the number of arcs only
$\lesssim 15\%$ \citep{flores00,meneghetti00}, though they may affect
the number of tangential and radial arcs differently. Finally,
deviations from spherical symmetry of lens halos change the number of
arcs \citep{bartelmann95b,molikawa99}, which will increase the number 
of tangential arcs and decrease the number of radial arcs \citep{molikawa01}. 
Therefore this effect is certainly important in discussing the number 
ratio $W$ as well as the total number of arcs.
We plan to incorporate these effects in a systematic and quantitative 
fashion which will be reported elsewhere.

\acknowledgments

We thank Takeshi Chiba, Takashi Hamana, and Ryuichi Takahashi for useful
discussions and Xiang-Ping Wu for his instructive comments.  We also
thank an anonymous referee for many useful comments which improved the
earlier manuscript. This research was supported in part by the
Grant-in-Aid by the Ministry of Education, Science, Sports and Culture
of Japan (07CE2002) to RESCEU.

\clearpage

\clearpage

\clearpage
\clearpage
\clearpage
\clearpage
\clearpage
\clearpage
\clearpage
\end{document}